\documentclass[aps,twocolumn,amssymb,floatfix,showpacs,preprintnumbers,amsmath,amssymb,superscriptaddress]{revtex4}
\usepackage{graphicx}
\relax

\begin{document}
\title{Length dependence of the resistance in graphite: Influence of ballistic transport}
\author{P. Esquinazi}\email{esquin@physik.uni-leipzig.de}
\author{J. Barzola-Quiquia}
\author{S. Dusari}
\affiliation{Division of Superconductivity and Magnetism, Institut
f\"ur Experimentelle Physik II, Universit\"{a}t Leipzig,
Linn\'{e}stra{\ss}e 5, D-04103 Leipzig, Germany}
\author{N. Garc\'ia}
\affiliation{Laboratorio de F\'isica de Sistemas Peque\~nos y
Nanotecnolog\'ia,
 Consejo Superior de Investigaciones Cient\'ificas, E-28006 Madrid, Spain}

\begin{abstract}
Using a linear array of voltage electrodes with a separation of several micrometers
on  a $20~$nm thick and 30~$\mu$m long
multigraphene sample we show that the measured resistance
does not follow the usual length dependence according to Ohm's law. The deviations can be quantitatively explained taking into account  Sharvin-Knudsen formula for ballistic transport. This allows us to obtain without free parameters the  mean free path of the carriers in the sample at different temperatures. In agreement with recently reported values obtained with a different experimental method, we obtain that the carrier mean free path is of the order of  $\sim 2~\mu$m with a mobility $\mu \sim 10^7~$cm$^{2}$V$^{-1}$s$^{-1}$. The results
indicate that the usual Ohm's law is not adequate to calculate the
absolute resistivity of mesoscopic graphite samples.
\end{abstract}
\maketitle

\section{Introduction}
One of the important parameters that determines the electronic transport in a material is the temperature and magnetic field dependent mean free path $\ell$. Its direct measurement, model and parameter-free independent, is however difficult and only in some special cases possible, like in materials with relatively large mean free path. Among those materials   are the ones, which show   ballistic transport, at least for a given device channel length like in some carbon nanotubes samples \cite{kon01,lia01,jav04}. The direct measurement of $\ell$ is possible, for example, with a scanning microscope through the scaling of the channel resistance \cite{par04} or with a multi terminal method \cite{gao05}.
In Ref.\onlinecite{pur07} the electron mean free path of a carbon nanotube was obtained from a scaling of the resistance with length,
getting  $\ell(300$K$) = 0.2 \ldots 0.8~\mu$m upon the nanotube sample.
Large values of $\ell $ were found also in suspended graphene reaching ballistic transport in the micrometer range at low temperatures \cite{bol08,du08}.

Another  method to obtain $\ell$ is the constriction method,
based on the
measurement of the longitudinal resistance $R$
 as a function of the width $W$ of a constriction located between the voltage electrodes \cite{gar08}.
When $\ell \gtrsim W$ the ballistic contribution overwhelms the diffusive ones
 allowing  to obtain $\ell(T)$  without the need of free
parameters or arbitrary assumptions. This method has been
used to obtain $\ell(T)$ and the mobility $\mu(T)$  in bulk highly oriented pyrolytic graphite (HOPG) samples \cite{gar08} as well as in some tens of nanometers thick and micrometers
large multigraphene samples \cite{dus11}.  The obtained results indicate that the mean free path of the carriers
within the graphite layers inside the graphite structure is indeed large, reaching the micrometer range even
at room temperature \cite{dus11}.  However, the constriction method needs to cut part of the sample, i.e.
a constriction of the size of the order of the mean free path has to be patterned in the sample middle without
affecting its internal structure. Although this is possible using a focused ion beam and a protecting film
\cite{barnano10}, one may still doubt whether the huge increase in resistance observed for constriction
widths of the order of several micrometers \cite{dus11} is intrinsic and not due to the influence of
the ion beam on the graphite sample. Therefore, it is necessary to obtain the mean free path and other
transport parameters like the mobility using more transparent and less invasive methods to check whether the
obtained mean free path and mobility in multigraphene are really as large as reported previously.

In this work we present a simple method to obtain $\ell$ in thin, mesoscopic graphite flakes based on the dependence of the longitudinal resistance with the distance between the voltage electrodes on the sample,
a method somehow related to those used in Refs.~\cite{gao05,pur07}. As we show in this work, due to the micrometer large mean free path of the carriers in multigraphene, the
ballistic contribution to the measured resistance becomes already appreciable for sample sizes several times larger than $\ell$. The  experimental approach presented in this work represents a more transparent alternative to the constriction method used in previous works and the obtained results support basically the result that $\ell(300$K$) \gtrsim 1~\mu$m in highly ordered thin graphite of mesoscopic size and of good quality.

\section{Experimental details}

\begin{figure}
\includegraphics[width=1\columnwidth]{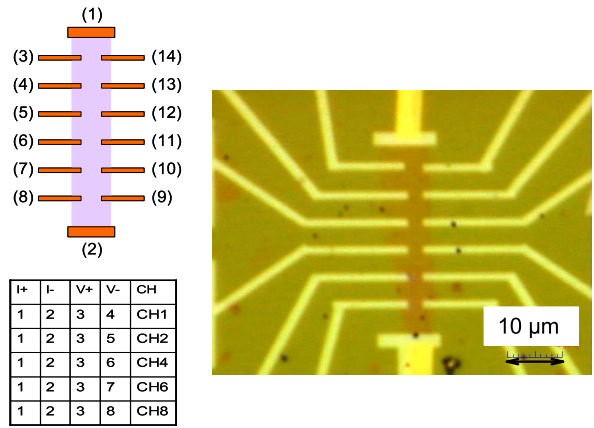}
\caption{The upper left sketch shows the current and voltage
electrodes configuration for the different channels. The middle
right picture is an optical microscope photo of the measured sample
with all its electrodes. The bottom left
table defines the channels with the corresponding configuration
for current and voltage electrodes.} \label{ske}
\end{figure}

A $\simeq 30~\mu$m long, $(5 \ldots 8.5) \pm 0.3~\mu m$ wide and $(20 \pm 2)$~nm thick
multigraphene sample was contacted with 14 electrodes  prepared by electron-beam lithography and Pd/Au deposition,
see Fig.~\ref{ske}. For the determination of the absolute value of the mean free path
the distance between electrodes is of importance; each electrode had a width of $(1.36 \pm 0.1) \mu$m.  Taking the distance between the middle point of the
electrodes we have then $L_0 = (4.8\pm 0.2)~\mu$m.
The multigraphene sample was prepared by a rubbing
and a ultrasound method described in Ref.~\onlinecite{bar08} and from a HOPG sample with $0.35^{\circ}$ rocking curve width. Micro-Raman measurements indicate that even samples of 10~nm thickness are of good quality without showing any contribution of the defect-related D-peak at 1350~cm$^{-1}$~\cite{gar11}.

With an AC bridge from Linear
Research (LR700) we measured the resistance $R(T,L)$ at different temperatures
$T$ and at different lengths $L$ between voltage electrodes. The
inset in Fig.~\ref{ske} shows the different configuration
channels. For example, channel~1 means input current at electrodes
1 and 2 and voltage measurements at electrodes 3 and 4. The
results we discuss below are independent whether we take the
electrodes array at the right or left of the sample, i.e. 3 and 4 or 14 and
13 (see inset in Fig.~\ref{ske}) indicating a homogeneous behavior
of the sample and the current distribution. The used input AC current was $I = 2~\mu$A in all
the measurements.

\begin{figure}
\includegraphics[width=1\columnwidth]{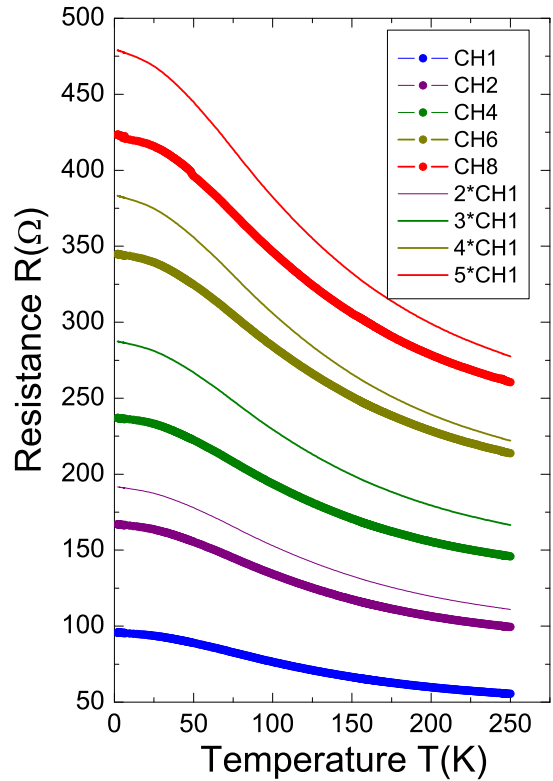}
\caption{Temperature dependence of the resistance of the
multigraphene sample at five different channels defined in the
inset of Fig.~2. The continuous lines represent the expected
resistance if it would be just only proportional to the length
between electrodes.} \label{TR}
\end{figure}

\section{Results}

\begin{figure}
\includegraphics[width=1\columnwidth]{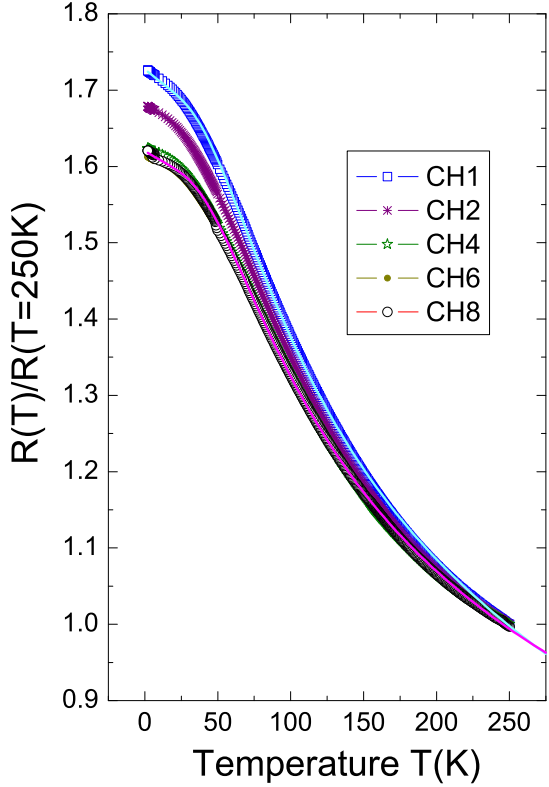}
\caption{Temperature dependence of the normalized resistance at
the different channels. The continuous lines are fits following
the parallel resistor model with a semiconducting and an interface
contribution. All curves are fitted with the same semiconducting energy gap of
$E_g = 350$[K]/$k_B$.} \label{TRN}
\end{figure}

Figures \ref{TR} and \ref{TRN} show the temperature dependence of the
absolute and normalized resistance of the sample at different
channels, respectively. In Fig.~\ref{TR} we also show the expected $R(T)$ of the channels CH2 to CH8 if the resistance would be just proportional to the distance between voltage electrodes, i.e. $R_{\mathrm{CH}_i} (T)= (L_i/L_0) R_{\mathrm{CH}_1}(T)$, with $L_i/L_0 = 2 \ldots 5$.
As one can clearly recognize in that figure, none of the measured curves follows the expected diffusive Ohmic behavior but are below the one expected. This disagreement is independent of the effective
value of $L_0$, i.e. whether we take it between the middle points of the electrode widths or
just between the nearest edges ($\simeq 3.5~\mu$m).

Taking into account that the width of the sample is not constant
but decreases with the channel number, see Fig.~\ref{ske}, the
expected $R_{\mathrm{CH}_i}(T)$ and according to Ohm's law should
be even larger, i.e. $R_{\mathrm{CH}_i} (T)= (L_i/L_0)(W_1/W_i)
R_{\mathrm{CH}_1}(T)$, where $W_1 = 8.3 \pm 0.2~\mu$m is the width
at channel 1 and $W_i$ an average width between the measured width
at electrode 3 ($W_1$) and the width $W_i^\ast$ at the
corresponding end electrode of the channel $i$. This width $W_i$
can be estimated either by the geometric mean or a simple average,
i.e. $W_i = \sqrt{W_1W_i^\ast}$ or $(W_1 + W_i^\ast) /2$, The
small differences between the two effective widths change only
slightly the absolute estimate of the mean free path and are not
relevant.

The failure of the diffusive Ohm's law to describe the data can be also recognized in Fig.~\ref{TRN}. In the scale of this figure the larger  the distance between electrodes the better is the normalization. As we will see below this is exactly what we expect if the ballistic transport plays a role in the
observed behavior. Note that the semiconducting-like behavior with a saturation at low temperatures
is observed at all distances between the voltage electrodes indicating that this dependence is intrinsic and not related to the sample size, supporting the conclusions of Ref.~\onlinecite{gar11}. In fact, using the
same parallel resistance model and the exponential temperature dependence appropriate for semiconductors as in Ref.~\onlinecite{gar11}, the obtained energy gap $E_g/k_B \simeq 350~$K is independent of the distance between electrodes.

\begin{figure}
\includegraphics[width=1.0\columnwidth]{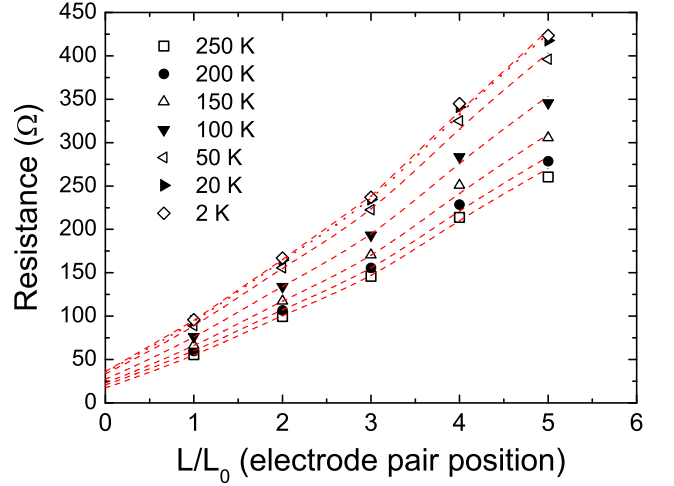}
\caption{Length dependence of the measured resistance. The unit
length distance $L_0$ between electrodes is for all electrodes the
same. The dashed and dotted (20~K) lines are obtained from the
fits to Eq.~(\ref{1b}) with only the ballistic resistance $R_0$ as
fitting parameter.}{\label{LR}}
\end{figure}

Figure~\ref{LR} shows the absolute
resistance as a function of the channel length at different
temperatures. The experimental points
appear to be consistently non-linear, curving up with the distance
to the first channel. The reason for the observed dependence is explained below
and allows us to obtain the carrier mean free path.

\begin{figure}
\includegraphics[width=1.0\columnwidth]{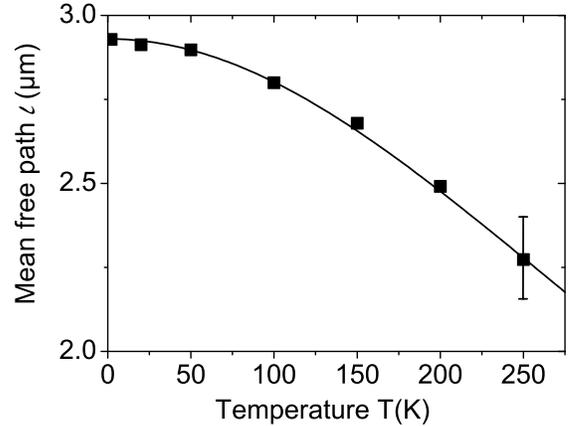}
\caption{Temperature dependence of the mean free path obtained from
Eq.~(\ref{ell}). The error bar is estimated through the error in the determination
of $R_0(T)$ from the fits of the data to Eq.~(\ref{1b}), see  Fig.~\protect\ref{LR}.
The continuous line follows the
equation $\ell(T) = ((2.93)^{-1} + ((6.4 \times 10^5)/T^2)^{-1})^{-1}$ ($T$ in K and $\ell$ in $\mu$m).}{\label{tele}}
\end{figure}

 According to
Sharvin-Knudsen formula and  Ohm's law, the resistance of the
sample in terms of geometrical parameters and the resistivity $\rho$,
is given by
\begin{equation}
R = \frac{{\rho \left( T \right)}}{{tW}}\left[ {\ell \left( T
\right) + L} \right]\,, \label{1}
\end{equation}
where  $\ell$ is the mean free path, $t$ the thickness, $W$
the width of the sample and $L$ the distance between the
voltage electrodes for each of the selected channels.
Defining the ballistic resistance   $R_{0} = R(L = 0)$ as the first term in Eq.~(\ref{1}),
 taking into account the effective channel width $W_i$, assuming homogeneous resistivity and
 thickness through the whole sample length, this equation can be rewritten as:
\begin{equation}
R_i(T) = R_{0}(T) + (R_1(T) - R_0(T)) \frac{W_1}{W_i} \frac{L_i}{L_0} \,, \label{1b}
\end{equation}
where the width at channel 1 is $W_1 = 8.3 \pm 0.2~\mu$m, $W_i$'s are
estimated as explained  above and $L_i / L_0 = 0 \ldots 5$.

With the measured widths the experimental data shown in Fig.~\ref{LR} can be very well
fitted using Eq.~(\ref{1b}) having  $R_0(T)$ as the only
free parameter. To obtain the mean free path from $R_0(T)$ we need to calculate the resistivity
$\rho(T)$. This last is  obtained from:
\begin{equation}
\rho(T) =\frac{tW_1}{L_0} (R_1(T) - R_0(T)) \,,
\label{rho}
\end{equation}
and therefore the mean free path is calculated from:
\begin{equation}
\ell(T) = \frac{R_0(T) L_0}{R_1(T)-R_0(T)}\,.
\label{ell}
\end{equation}
Using Eqs.~(\ref{rho}) we obtain $\rho(2$K$) = (208 \pm 40)~\mu \Omega$cm, where the error is due to
the errors in the geometry parameters and $R_0(T)$.
The mean free path obtained from Eq.~(\ref{ell}) and as a function of the temperature is shown in Fig.~\ref{tele}.
Note that the obtained mean free path is independent of the values of the
width and thickness of the sample and the statistical error is that of $R_0(T)$
from the fits in Fig.~\ref{LR}.  Within experimental error the temperature
dependence follows a simple parallel resistance model, i.e.
$\ell(T) = ((2.93)^{-1} + ((6.4 \times 10^5)/T^2)^{-1})^{-1}$ ($T$ in K and $\ell$ in $\mu$m),
with a temperature dependent term given by $T^{-2}$, similar to that found in HOPG bulk samples\cite{gar08}. This dependence suggests that  electron-electron interaction can be the main temperature dependent scattering process.

The error in the absolute value of the mean free path obtained in this work
resides mainly in the uncertainty of $L_0$.
One may speculate that the Pd/Au deposited electrodes are invasive and may affect the
current distribution and in this case $L_0 \sim (4.8 - 1.36)~\mu$m.
On the other hand the electrodes may affect only the upper most
graphene layer from the $\sim 60$ graphene layers inside the rest of the sample. Also,
the effective resistance of the polycrystalline Pd film electrode contacting the upper most
graphene layer along its $\simeq 1.4~\mu$m width is not necessarily smaller than the
mainly ballistic resistance of the graphene layer. Therefore, taking the electrode width as the
maximum uncertainty in $L_0$ we estimate $\sim 30\%$ as the largest
absolute error in the mean free path.

With the knowledge of the mean free path and the resistivity of the sample we can calculate the Fermi wavelength $\lambda_F$ and the mobility $\mu$  of the carriers using the equations $\lambda_F = 2 \pi e^2 \rho \ell / h a$ and $\mu = e \lambda_F \ell / h$ where $a$ is the distance between graphene planes in Bernal graphite and the other parameters are the usual natural constants. At 2~K (250~K)  we obtain $\lambda_F = 4.5 (2.3)~\mu$m and $\mu = 3 \times 10^7 (1.4 \times 10^7)~$cm$^2$V$^{-1}$s$^{-1}$ with a maximum error of 50\% in the absolute values. \\

\section{Conclusion}
The aim of this experimental work was to show in a single multigraphene sample of large enough length that the resistance does not follow the Ohm law. It does not increase strictly proportional to the distance between voltage contacts, as one would expect from the usual diffusive Ohm law. Instead, we  show experimentally that a length independent contribution to the resistance due to the ballistic transport of the carries needs to be considered. From our measurements and with the help of the Sharvin-Knudsen formula for ballistic transport we are able to obtain the mean free path of the carriers at different temperatures and without free parameters. The obtained results support previous results and indicate that the graphene layers within the graphite structure have micrometer  large mean free path and Fermi wavelength and mobility  $ \gtrsim 10^7~$cm$^2$V$^{-1}$s$^{-1}$ at 300~K.  We also conclude that for multigraphene samples of good quality and of  size in the micrometer range, a significant error in the estimate of the resistivity is done  if the ballistic contribution is not taken into account. Clearly, the results indicate that the
Boltzmann-Drude approach to obtain electronic parameters from transport measurements is not adequate.

This work was  supported by  the DFG under ES 86/16-1.
S.D. is supported by  the Graduate School of
Natural Sciences ``BuildMona".


\end{document}